\documentclass[reprint,
 amsmath,amssymb,
 aps,
pra,
titlepage,
floatfix,
superscriptaddress
]{revtex4-1}

\usepackage{graphicx}
\usepackage{dcolumn}
\usepackage{bm}
\usepackage{xcolor}
\usepackage{siunitx}

\newcommand{\ket}[1]{\ensuremath{\left|#1\right>}}

\newcommand{\yb}{$^{171}$Yb$^+$ }

\newcommand{\sfo}{$^2$S$_{1/2}{\left|F=1\right>}$ }

\newcommand{\pfz}{$^2$P$_{1/2}{\left|F=0\right>}$ }
\newcommand{\pfo}{$^2$P$_{1/2}{\left|F=1\right>}$ }

\begin{document}

\preprint{APS/123-QED}

\title{
High-Speed Low-Crosstalk Detection of a \yb Qubit using Superconducting Nanowire Single Photon Detectors
}

\author{Stephen Crain}
\author{Clinton Cahall}
\author{Geert Vrijsen}
\affiliation{%
 Fitzpatrick Institute for Photonics, Electrical and Computer Engineering Department, Duke University, Durham, North Carolina 27708, USA}%

\author{Emma E. Wollman}
\author{Matthew D. Shaw}
\affiliation{%
	Jet Propulsion Laboratory, California Institute of Technology, 4800 Oak Grove Dr., Pasadena, California 91109, USA}%

\author{Varun B. Verma}
\author{Sae Woo Nam}
\affiliation{%
	National Institute of Standards and Technology, 325 Broadway, Boulder, CO 80305, USA}%
    
\author{Jungsang Kim}%
\email{E-mail: jungsang@duke.edu}
\affiliation{%
 Fitzpatrick Institute for Photonics, Electrical and Computer Engineering Department, Duke University, Durham, North Carolina 27708, USA}%
 \affiliation{%
 IonQ, Inc, 4505 Campus Drive, College Park, MD 20740, USA}%

\date{\today}

\begin{abstract}
Qubits used in quantum computing suffer from errors, either from the qubit interacting with the environment, or from imperfect quantum logic gates. Effective quantum error correcting codes require a high fidelity readout of ancilla qubits from which the error syndrome can be determined without affecting data qubits. Here, we present a detection scheme for \yb qubits, where we use superconducting nanowire single photon detectors and utilize photon time-of-arrival statistics to improve the fidelity and speed. Qubit shuttling allows for creating a separate detection region where an ancilla qubit can be measured without disrupting a data qubit. We achieve an average qubit state detection time of \SI{11}{\us} with a fidelity of \SI{99.931(6)}{\percent}. The detection crosstalk error, defined as the probability that the data qubit coherence is lost due to the process of detecting an ancilla qubit, is reduced to $\sim$2$\times10^{-5}$ by creating a separation of \SI{370}{\um} between them.


\begin{center}
\mbox{}
\vfill{}
Copyright 2018. All rights reserved.
\end{center}
\clearpage
\end{abstract}

\maketitle

Trapped ions have proven to be a highly effective implementation platform for quantum computing~\cite{Wineland1998,Kim2013,Monroe2016}, where the basic operations have been fully demonstrated to initialize and detect the qubit states, maintain coherence while the ions represent the qubits, and perform a universal set of quantum logic gates with high fidelity~\cite{Divincenzo2000}. For the \yb qubit used in our experiments, the qubit readout is performed by state dependent florescence using a cycling transition between one of the qubit states and an excited level of the atomic ion~\cite{Nagourney, Sauter, Bergquist}. The measurement time and fidelity is driven by the collection and detection efficiency of the scattered photons and background levels at the photon detector~\cite{Myerson2008,Noek2013}. Work has been done to improve the collection of the emitted photons from ions using various integrated optical strategies including standard high numerical aperture (NA) optics~\cite{Noek2013, Brady2011}, reflective mirrors~\cite{Shu2010,Maiwald2012}, Fresnel lenses~\cite{Streed2011}, integrated fiber optics~\cite{VanDevender2010}, and optical cavities~\cite{Kim2011_2,Sterk2012}.

Previous work relied on photomultiplier tubes (PMTs) and charge-coupled devices (CCDs) for photon detection which have limited quantum efficiencies of 20-30$\%$ near 369.5\,nm. In this work, we greatly improve the overall qubit state detection efficiency and speed by using superconducting nanowire single photon detectors (SNSPDs) customized for this application. SNSPDs have become the ubiquitous technology for photon counting applications because of their nearly ideal performance metrics. These detectors have high detection efficiency, low dark count rates, and high maximum count rates~\cite{dauler2014}. These superb performance characteristics compared to other single-photon detectors have made possible recent experimental demonstrations utilizing single photons, such as a loophole-free Bell inequality test~\cite{shalm2015strong}, Lunar Laser Communication Demonstration~\cite{boroson2014overview}, and high-rate quantum key distribution~\cite{islam2017provably}. 

In most atomic qubits where resonant fluorescence is used for state detection, the scattered photons from the pump beam or the atoms can cause decoherence in nearby qubits, by the same state-dependent fluorescence process used for qubit state detection.  Certain quantum circuits, such as quantum error correction, require that ancilla qubits be measured in the middle of the algorithm \cite{Steane2007,Tomita2014}. It is critical to preserve the memory of the data qubits while this resonant, destructive detection process is performed on the ancilla qubits. One approach is to transfer the ancilla qubit into a different ion species and detect it using light at a different wavelength, so the pump beam does not affect the data qubits~\cite{SchmidtScience2005,HumePRL2007}. Another approach is to leverage the segmented control electrodes of microfabricated surface traps to create multiple trapping zones~\cite{slusher,Amini}, and spatially separate  the ancilla qubits from the data qubits for the state detection process~\cite{Kielpinski2002}. We quantitatively measure the decoherence of the data qubits due to the qubit detection process when the ancilla qubit is spatially separated from the data qubits.

In this work, we achieve an average qubit state detection time of \SI{11}{\us} with a fidelity of \SI{99.931(6)}{\percent} using a high numerical aperture lens for photon collection and an SNSPD for photon detection. The high detection efficiency and low dark count rate of the SNSPD allows us to wait for only a single detection event for the state detection of the \ket{1} state. For an experiment with both a data and ancilla qubit, the detection crosstalk error is defined as the probability that the data qubit coherence is lost due to the process of detecting an ancilla qubit. When the ancilla qubit is shuttled \SI{370}{\um} away from the data qubit, the detection crosstalk error is reduced to $\sim$2$\times10^{-5}$.

\begin{figure}[tpb]
	\begin{center}
		\includegraphics[width=\columnwidth]{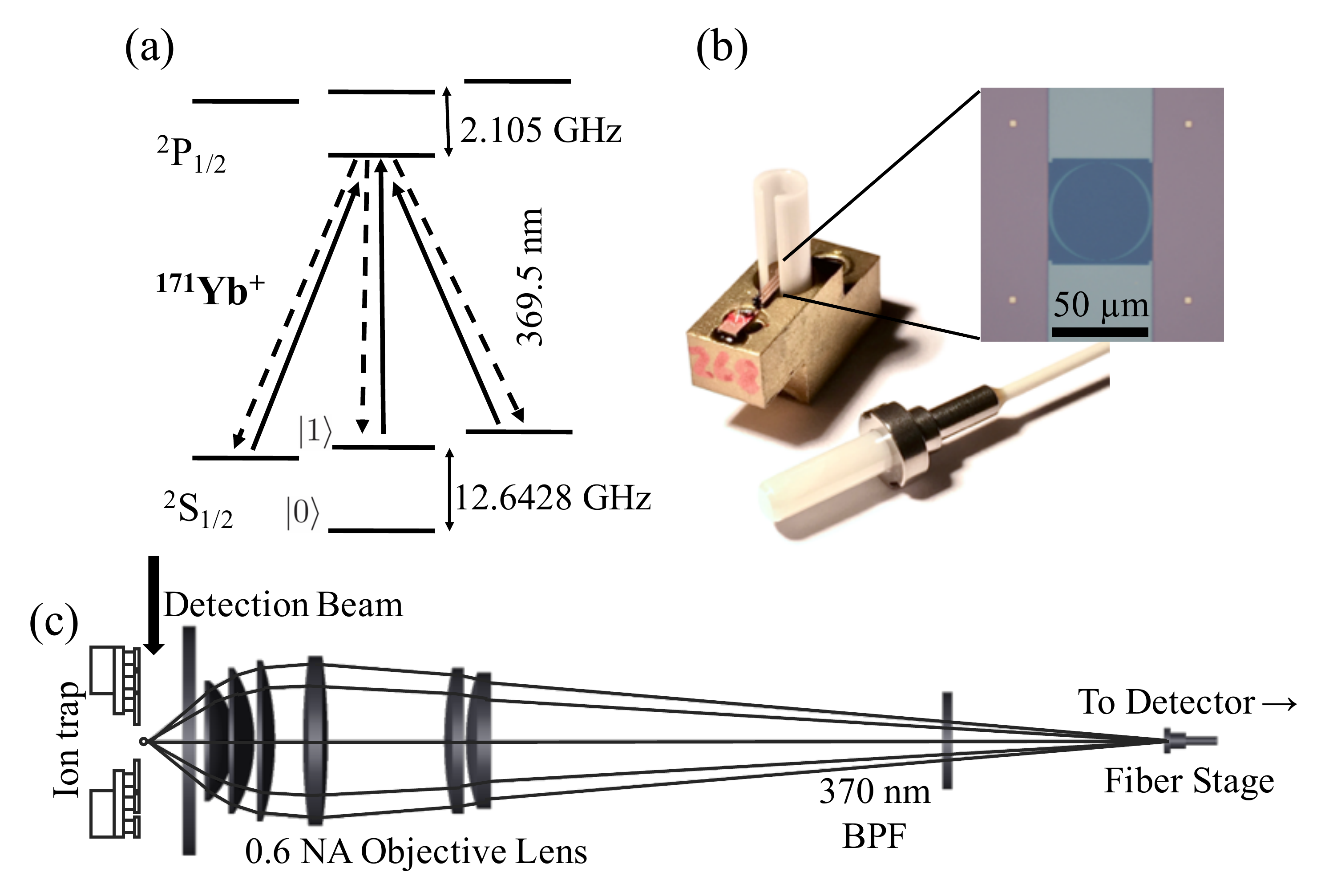}
		\caption[]{\textbf{Experimental System} (a) Diagram of the relevant energy levels of the \yb qubit. (b) Image of the self-aligning fiber package for the SNSPD with a standard fiber ferrule for scale. The inset shows an SEM image of the SNSPD device with the active area in blue. (c) Photons scattered from a single trapped ion are coupled into a multi-mode fiber through a 0.6 NA imaging lens and a bandpass filter (not to scale). }
		\label{fig:experimentalSetup}
	\end{center}
\end{figure}


\section{Results}

\subsection*{High Speed and High Fidelity State Detection}
The relevant energy levels of the \yb ion are shown in Figure \ref{fig:experimentalSetup}(a). When the ion is in the \ket{1} state, the resonant \SI{369.5}{\nano\meter} laser beam pumps the ion to the $^2$P$_{1/2}$ excited state, where it will spontaneously emit a photon as it transitions back to the three ``bright'' states. The detection beam is not resonant with any transitions for an ion in the \ket{0} state, so the ion will scatter a negligible number of photons from this ``dark'' state. For this experiment, we trap a single ytterbium ion \SI{70}{\um} above the surface of a microfabricated radio frequency (RF) Paul trap from Sandia National Laboratories. A detection beam with a \SI{15}{\micro\meter} waist propagating across the surface of the trap is directed onto the ion. The detection beam is delivered to the trapping location via a single mode fiber and an achromat focusing lens. A custom imaging lens with numerical aperture of 0.6 (Photon Gear, Inc) is used to collect \SI{10}{\percent} of the photons scattered from the ion. The photons pass through a \SI{6}{\nm} bandpass filter and are coupled into a multi-mode fiber with a core diameter of \SI{50}{\um}, which also acts as a spatial filter for unwanted scatter from the detection beam. The fiber enters a cryostat where it is coupled to an SNSPD in a self-aligning package, shown in Figure \ref{fig:experimentalSetup}(b). Recent work has shown progress towards integrating the SNSPD into the ion trap structure itself \cite{Slichter17}. A diagram of the experimental setup is shown in Figure \ref{fig:experimentalSetup}(c). The overall detection efficiency of a photon scattered from the ion to be detected by the photon counting detector is defined as $\varepsilon_{sys}$ (see Eq.~(\ref{eq:sysEff}) in Methods).

The state detection process of \yb is dependent on three scattering rates: the scattering rate of the bright \ket{1} state ($R_o$), the bright pumping rate ($R_b$), and the dark pumping rate ($R_d$) \cite[]{Noek2013}. The bright pumping rate is the rate at which the qubit starts in the dark \ket{0} state and off-resonantly scatters into the bright \ket{1} state. Similarly, the dark pumping rate is the probability of the qubit starting in the bright \ket{1} state and off-resonantly scattering into the dark \ket{0} state. These scattering rates can be experimentally measured by preparing the ion in the \ket{1} state and varying the detection time over which photons are collected \cite[]{Noek2013}. With a sufficiently high photon detection rate $\varepsilon_{sys} R_o$, a low background count rate ($R_{bg}$), and $R_b$ slow compared to the detection interval, the state discriminator threshold can be set to zero.  If a single photon is detected during the detection interval, the qubit is determined to be in the \ket{1} state. The errors associated with this state detection method can be classified as either a detection error of the bright state or a detection error of the dark state. 

The probability of detecting $n$ photons in a detection interval $t$ with an initial photon collection rate given by $R_1$, and a state transition rate $R_T$ that changes the ion from its initial state to one that has a photon collection rate $R_2$, is given by
\begin{multline}P(n;t,R_1,R_2,R_T) = \\ \sum_{k=0}^{n}\int_{0}^{t}d\tau P_p(k;R_1\tau)\tilde{p}(\tau,R_T)P_p(n - k; R_2(t - \tau)),\label{eq:stateDetectIntegral}\end{multline}
where $P_p(n;\overline{n})$ is the Poissonian probability of detecting $n$ photons given $\overline{n}$ expected photons, and $\tilde{p}(t,R) = \frac{\mbox{d}}{\mbox{dt}}\left(1-e^{-Rt}\right) = Re^{-Rt}$ is the probability density for the photon collection rate to change at time $\tau$, as a Poissonian event with rate $R$.

Substituting the appropriate rates into Eq.~(\ref{eq:stateDetectIntegral}), the probability to detect zero photons for a given detection time ($t$) from an ion that is initialized to the \ket{0} state is

\begin{multline}
P_{t,d}(n=0) = \frac{R_b}{\varepsilon_{sys} R_o - R_b}e^{-R_{bg}t}\left[e^{-R_{b}t} - e^{-\varepsilon_{sys} R_{o}t}\right]\\+e^{-R_{b}t}e^{-R_{bg}t},
\label{eq:darkProbDetectZero}
\end{multline}
where $n$ is the number of detected photons in the time interval [0,$t$]. Similarly, the probability to detect zero photons from an ion that was initialized to the \ket{1} state for a given detection time ($t$) is
\begin{multline}P_{t,b}(n=0) = \frac{R_d}{\varepsilon_{sys} R_o + R_d}e^{-R_{bg}t}\left[1 - e^{-(\varepsilon_{sys} R_o + R_d)t}\right]\\+e^{-R_{d}t}e^{-(R_{o}+R_{bg})t}
\label{eq:brightProbDetectZero}
\end{multline}
The state detection error for the \ket{0} state is $1 - P_{t,d}$, and the state detection error for the \ket{1} state is $P_{t,b}$. 

The \yb ion is prepared in the \ket{0} state by applying a field resonant with the \sfo $\rightarrow$ \pfo transition. The light resonant with this transition is generated by adding sidebands to the 369.5\,nm cooling beam with a \SI{2.1}{\GHz} EOM. We estimate the error in the \ket{0} state preparation from imperfect pumping to be $\sim$10$^{-6}$, which is much less than other sources of error. In order to prepare the ion in the \ket{1} state, a microwave field resonant with the hyperfine transition is applied to rotate the ion to the \ket{1} state from the \ket{0} state. The fidelity of the gate to rotate the qubit to the \ket{1} state is measured by performing gate set tomography (GST), which is independent of state preparation and measurement \cite[]{Greenbaum, Blume2013, Blume2017}. For a target gate rotation of $0.5\pi$, the calculated rotation angle error in our experiment is $0.02(1)\%$.  

To measure the state detection fidelity of the \ket{0} and \ket{1} states, the detection beam is turned on for a set time, $\tau_d$ (\SI{500}{\us}), and the total number of photons detected by the SNSPD as well as each individual photon's arrival time with respect to the beginning of the detection interval is recorded by a field-programmable gate array (FPGA). The FPGA uses a \SI{200}{\MHz} clock to record the arrival times, resulting in a \SI{5}{\ns} timing resolution.  After the data is recorded, the arrival time of the first photon in each of the 100,000 experiments is extracted.  For the \ket{0} state, the state detection error for a particular $\tau_d$ is determined by the fraction of events where at least one photon arrives within the interval.  For the \ket{1} state, the state detection error corresponds to the fraction of events where no photons are detected within $\tau_d$.  

\begin{figure}[tpb]
	\begin{center}
		\includegraphics[width=0.95\columnwidth]{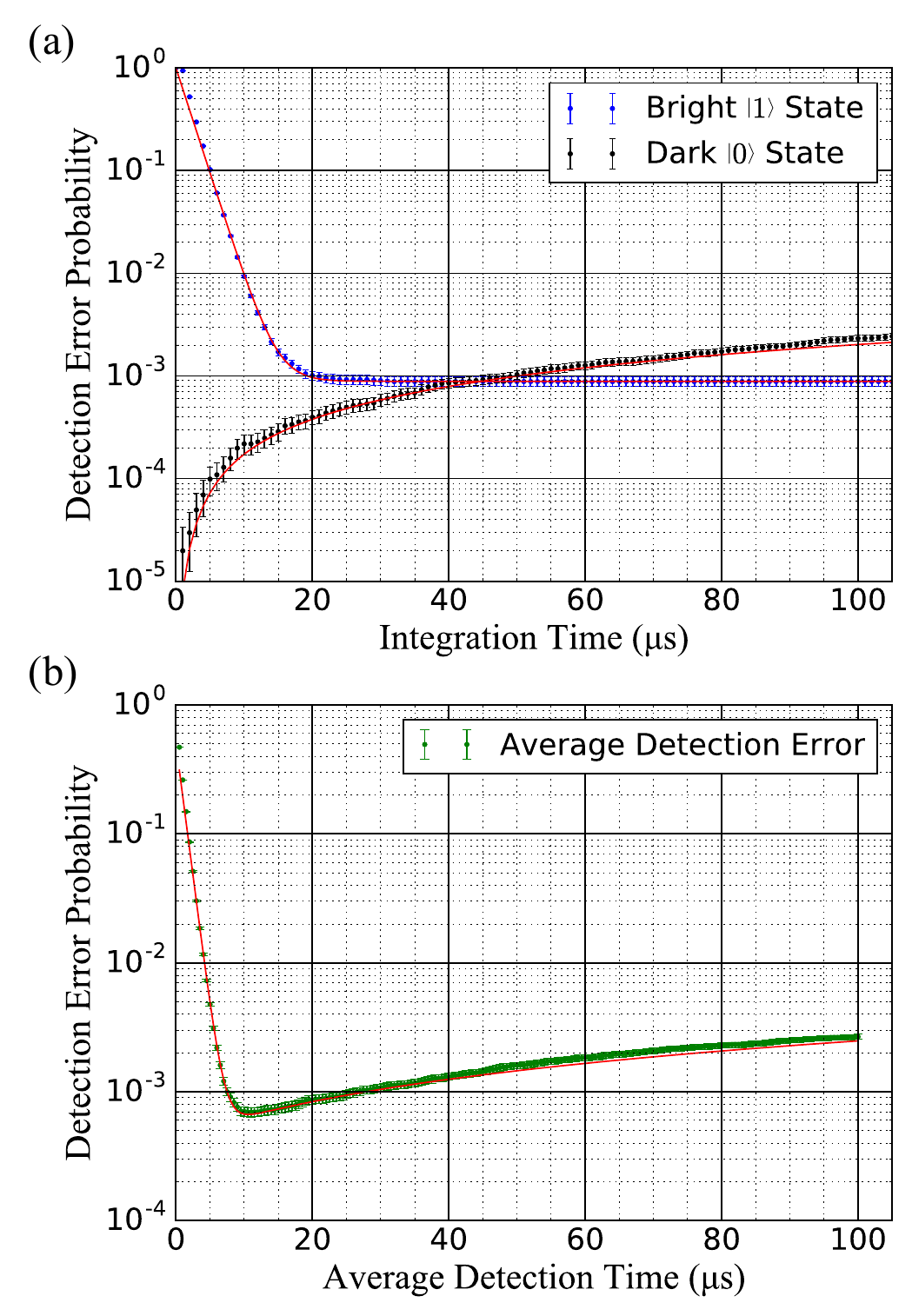}
		\caption[]{\textbf{Detection Error Probabilities} (a) The detection error probability as a function of integration time ($t_{d}$) when the ion prepared in the \ket{0} (dark) and \ket{1} (bright) state (100,000 experiments for each case) and (b) the average state detection error probability as a function of average detection time (the detection process stops upon the detection of a single photon). The solid lines show the analytic solution using measured dark pumping, bright pumping, total photon detection rates, and state preparation errors. The error bars indicate the standard error of the data.}
		\label{fig:avgDetectTime}
	\end{center}
\end{figure}

Figure \ref{fig:avgDetectTime}a shows the dark and bright state detection error probabilities as a function of $\tau_d$, with the corresponding analytical solution with no fit parameters. The dark state detection fidelity is limited by both the background count rate (\SI{4.2(1)}{cps}) and bright pumping rate (\SI{16.4(5)}{\Hz}), while the bright state detection fidelity is limited by the dark pumping rate (\SI{341(13)}{\Hz}) given our overall photon detection rate (\SI{472(14)}{kcps}). In order to discriminate a bright state from a dark state, one must choose a sufficiently long $\tau_d$ to minimize the state detection error probability for the \ket{1} state. When no photons arrive during $\tau_d$, the qubit is determined to be in the \ket{0} state. One can determine the qubit to be in the \ket{1} state upon detection of the first photon and complete the detection process, without waiting for the entire duration $\tau_d$ \cite{LangerPHD,HumePRL2007}. This detection process reduces the average qubit state detection time to be shorter than $\tau_d$, by up to a factor of $\sim 2$.
Figure \ref{fig:avgDetectTime}b shows the detection error probability of 200,000 experiments as a function of the average detection time. For a detection beam intensity of \SI{56.2}{mW cm^{-2}}, the average detection time is \SI{11}{\us} with 99.931(6)\% state detection fidelity.

\subsection*{State Detection Crosstalk}
To assess the impact of an ancilla qubit state detection on the coherence of a nearby data qubit, we need to split a chain of two qubits and shuttle one some distance away from the other~\cite{Amini2010}. For this experiment, digital to analog converters (DAC8734 from Texas Instruments) are used to generate the DC trapping voltages of $\pm$10\,V with 16 bit resolution (corresponding to \SI{300}{\uV} steps)~\cite{Mount2016}. Up to 100 unique voltages are asynchronously updated in real time by an FPGA (Opal Kelly XEM6010) at a maximum update rate of \SI{430}{kHz} (\SI{2.32}{\micro\second} per step). Shuttling solutions are pre-calculated to move an ion in \SI{5}{\micro\meter} steps along the entire 3 mm trapping zone, including solutions to split and merge a pair of ions.

\begin{figure}[tpb]
	\begin{center}
		\includegraphics[width=\columnwidth]{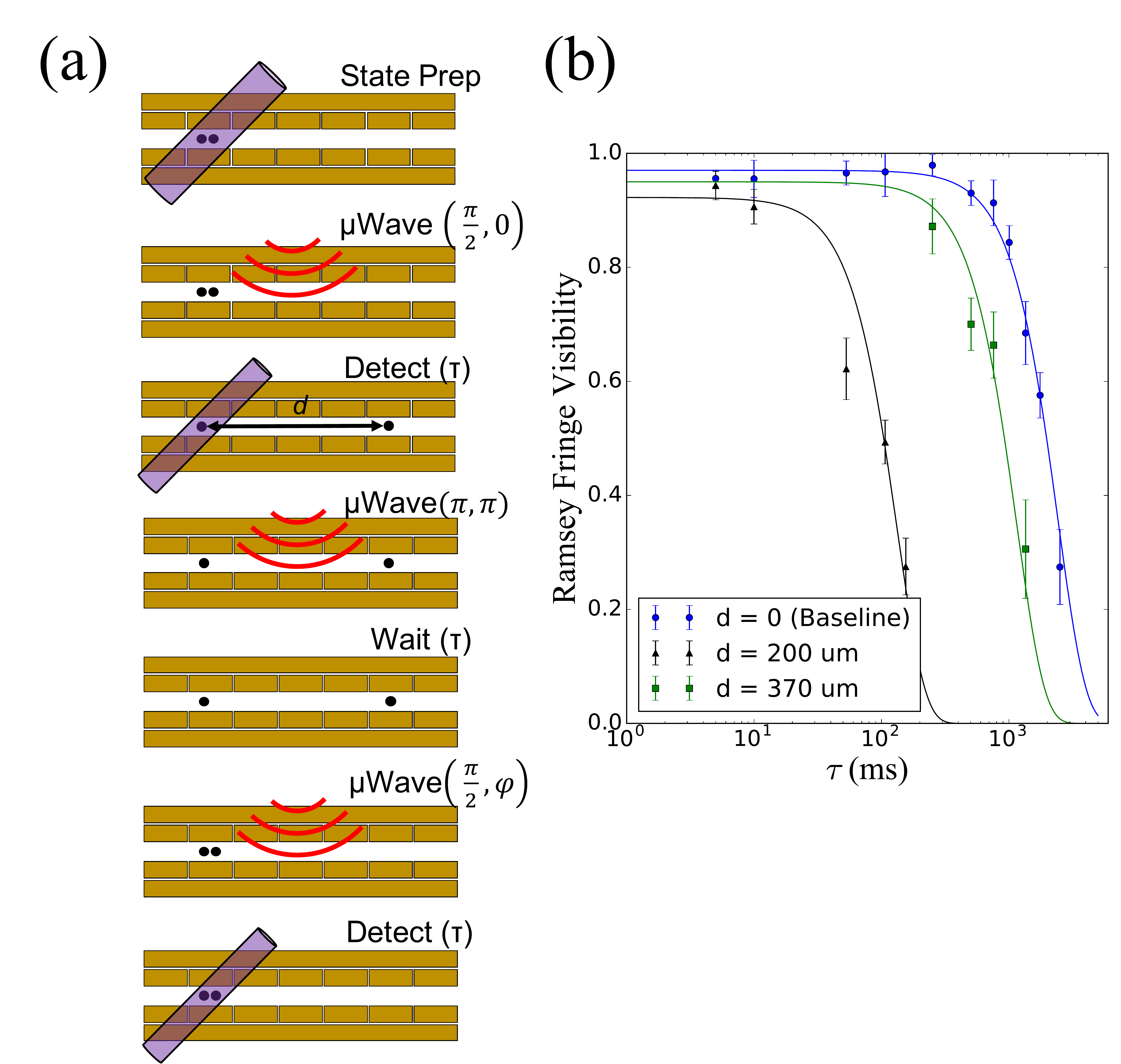}
		\caption[]{\textbf{Decoherence Due to State Detection Measurement} (a) Schematic of the experimental sequence (not to scale). (b) Results of the Ramsey experiment with three different shuttling parameters. The coherence time is extracted from the fit of fringe visibility as a function of time between the microwave pulses. The error bars indicate the standard error of the data.}
		\label{fig:shuttleRamsey}
	\end{center}
\end{figure}

Figure \ref{fig:shuttleRamsey}a shows the experimental sequence for a Ramsey-type experiment with a spin-echo to measure the coherence time of the data qubit, for various distances from the detected ancilla qubit. The ions are first Doppler cooled and then prepared in the \ket{0} state. A global microwave field is then used to perform a $\pi/2$ gate on the two qubits. The two qubits are split and shuttled a distance $d$ away from each other. A resonant \SI{369.5}{\nm} detection beam is applied to the ancilla qubit for a variable amount of time, $\tau$. A spin-echo pulse $R(\pi,\pi)$, followed by a waiting period of $\tau$, is applied to remove any constant frequency offset between the qubit and microwave source. The qubits are then merged back into a single trapping zone, followed by global microwave $\pi/2$ analysis pulse with a varying phase, $\phi$. A final detection pulse is used to read the state of the data qubit.

The results of the Ramsey experiment are shown in Figure \ref{fig:shuttleRamsey}b for various shuttling distances. To determine the coherence time, the Ramsey fringe visibility is fit to the function $Ae^{-\tau^2/\alpha^2}$ where $\alpha$ is the coherence time. As a baseline, the coherence time of the qubit is measured without shuttling or the detecion beam turned on, and is calculated to be $1,716\pm80$\,ms. For a \SI{200}{\um} and \SI{370}{\um} shuttling lengths, the coherence time is measured to be $94\pm5$\,ms and $814\pm77$\,ms, respectively. When the data qubit is closer to the detection region, it is exposed to more of the resonant detection beam causing more dephasing, which is evident in the fringe contrast reduction. The total round-trip shuttling time for these two distances is \SI{92.8}{\micro\second} and \SI{171.68}{\micro\second}, respectively. Using the average detection time of \SI{11}{\us} from the state detection experiment, we conclude that the data qubit will dechohere after approximately $6\times10^{4}$ measurements when the data qubit is shuttled \SI{370}{\um} away from the ancilla qubit. This corresponds to a measurement crosstalk of $\sim$2$\times10^{-5}$, defined as the probability that the data qubit will lose its coherence as the ancilla qubit is measured. We note that this limitation is set by the undesirable scattering of the pump beam inside the vacuum chamber, and this can further be reduced by improving the exit path of the pump beam from the vacuum chamber. The expected crosstalk from the emission of a photon from the ancilla qubit being absorbed by the data qubit \SI{370}{\um} away is $\mathcal{O}(10^{-6})$. 


\section{Discussion}
In this work, we have demonstrated that the enhanced detection efficiency of SNSPDs can increase the overall detection fidelity and significantly improve the state detection time of a trapped ion qubit. Due to the high detection efficiency and low dark count rate of the SNSPD, state detection of the \ket{1} state only relies on the detection event of a single photon, reducing the average detection time. When the background counts due to unwanted scattered photons is fully eliminated, we expect the average detection fidelity of the qubit to be limited by the bright and dark pumping rates due to the atomic structure of the ion. The fundamental fidelity limit with zero background counts is expected to be 99.941$\%$ at the photon detection efficiency levels achieved in our experiment, compared to the 99.931(6)$\%$ fidelity demonstrated with the experimental background levels of 4.2(1) cps in our setup. Further enhancement to the fidelity can also be achieved using a shelving technique \cite{Myerson2008,Harty2014} and other methods \cite{HumePRL2007,Schaetz2005}. The reduced detection time leads to dramatically reduced crosstalk that causes a nearby data qubit to decohere while ancilla qubits are being measured.

\section{Methods}
\subsection*{SNSPD Specifications}

An SNSPD consists of a current-biased superconducting nanowire with typical cross-sections of $\sim$5$\times$\SI{100}{\nm}. A load resistor is connected in parallel with the detector. When a photon is absorbed in a current carrying nanowire it locally disrupts superconductivity and creates a hot-spot through heat deposition and breaking of cooper pairs to create non-equilibrium quasiparticles. Joule heating and quasiparticle diffusion cause the hot-spot to grow until a section of the nanowire turns resistive~\cite{semenov2001quantum,renema2014experimental} The large resistance due to the hot-spot diverts current out of the detector and in to the load resistor, typically the \SI{50}{\Omega} input of a low-noise amplifier. The resulting voltage across the load resistor is amplified and recorded.

The SNSPD detectors we use are made from the amorphous superconductor molybdenum silicide (MoSi) and are optimized for detection at \SI{369.5}{\nm}. They are patterned in a meander with a diameter of $\sim$\SI{56}{\um} to overlap with the \SI{50}{\um} core of the multi-mode fiber used to deliver the collected photons to the detector. The operating temperature of the detectors is \SI{850}{mK}. Our cryostat system consists of a closed-cycle, Gifford-McMahon (GM) cyrocooler and a helium-4 sorption fridge. This system is designed to support the operation of MoSi and other amorphous SNSPDs. We measure detection pulses using low-noise cryogenic amplifiers~\cite[]{Cahall2018} and low-noise room temperature amplifiers with a bandwidth of \SI{500}{\MHz} (Mini-Circuits ZFL-500LN). The background count rate of the system is measured to be \SI{4.2(1)}\,counts per second (cps), due solely to the scattered photons from the detection beam reaching the detector (\SI{0.075}{cps.cm^2 mW^{-1}}). The measured intrinsic dark count rate in similar detectors was measured to be \SI{<0.001}{cps} \cite{Wollman17}. The value for the hot-spot resistance $R_{det}$ is on the order of \numrange[range-phrase = --]{1}{10}\si{\kohm}.  

\subsection*{SNSPD Detector Efficiency Calibration}
The detector efficiency of the SNSPD is calibrated by measuring the total photon detection efficiency of the system using the $^{174}$Yb$^+$ isotope. The system photon detection efficiency is first measured with a calibrated PMT. The $^{174}$Yb$^+$ isotope can be modeled as a simple two-level system, where the on-resonance detection rate of the scattered photons as a function of pump laser intensity is given by

\begin{equation}
R_{^{174}Yb} (I) = \varepsilon_{sys} \left(\frac{\Gamma}{2}\right) \left(\frac{I/I_{sat}}{1 + I/I_{sat}}\right),
\label{eq:174ScatterRate}
\end{equation}

where $\varepsilon_{sys}$ is the total system photon detection efficiency, $I_{sat} =$ \SI{51}{mW cm^{-2}}, $\Gamma =$ \num{2\pi x 19.6}\si{MHz}. The intensity of the pump laser is defined to be $I = cn\epsilon_{0}|E|^{2}/2$, where $c$ is the speed of light, $n$ is the refractive index of the medium, $\epsilon_{0}$ is the vacuum permittivity, and $|E|$ is the amplitude of the electric field. By varying the intensity of the pump beam and measuring the rate of detected photons, the total system photon detection efficiency $\varepsilon_{sys}$ can be calculated by fitting the data to Eq.(\ref{eq:174ScatterRate}). 

\begin{figure}[tpb]
	\includegraphics[width=.9\columnwidth]{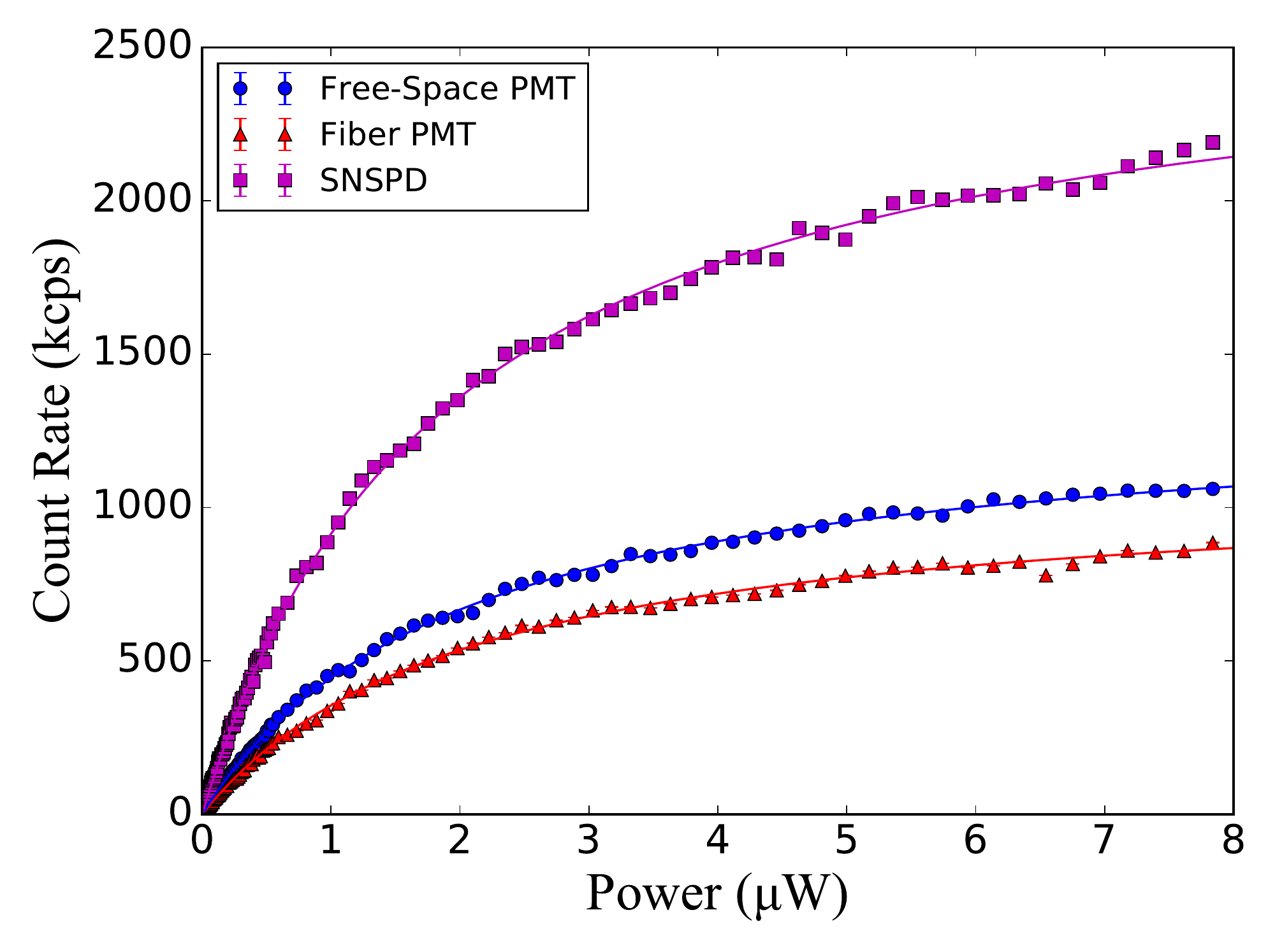}
	\caption[]{\textbf{Detector Efficiency Calibration} (Color online) The detection rate of the scattered photons from a single $^{174}$Yb$^{+}$ ion (measured in thousand counts per second, or kcps) as a function of pump beam power for various detection methods. The fit curve is scatter model of a two-level system. The statistical error bars are smaller than the data points.}
    \label{fig:collectionEfficiencies}
\end{figure}

Figure \ref{fig:collectionEfficiencies} shows the photon detection rate as a function of the power of the pumping beam using various detection methods. Each point is the average number of photons detected in 300 experiments with a \SI{0.5}{\ms} detection interval, converted to rate measured in counts per second (cps). Free space PMT (blue data) means that the photons collected by the high NA lens are detected directly by a PMT (Ultra Bialkali PMT, Hamamatsu). Fiber PMT (red data) means that the photons collected by the high NA lens are coupled into a multi-mode fiber, and detected by a nominally identical PMT on the other side of the fiber. SNSPD (purple data) means the multi-mode fiber delivers the photons to SNSPDs in the cryostat. The total system detection efficiency is further broken down as

\begin{equation}
\varepsilon_{sys} = \varepsilon_{PG} \varepsilon_{FC} \varepsilon_{fiber} \varepsilon_{det},
\label{eq:sysEff}
\end{equation}

where $\varepsilon_{PG}$ is the collection percentage of the 0.6 NA lens (\SI{10}{\percent}), $\varepsilon_{FC}$ is the fiber coupling percentage, $\varepsilon_{fiber}$ is the transmission of the fiber and connector efficiency (independently measured to be \SI{73.1(8)}{\percent}), and $\varepsilon_{det}$ is the detection efficiency of the detector device used. The fiber coupling percentage (\SI{81.8(5)}{\percent}) is calculated by comparing the total system efficiencies between the free space PMT and fiber PMT detection schemes. Based on the ratio of the fiber PMT and fiber SNSPD detection efficiencies, we determine the detection efficiency of the SNSPD device to be $79\% \pm 1.2\%$, given that that PMT quantum efficiency is nominally 32\% at 370 nm.

\begin{table}[tpb]
\renewcommand{\arraystretch}{1.25}
\caption[Summary of the total photon collection efficiencies using various detection methods.]{\textbf{Collection Efficiencies} Summary of the collection efficiencies of various detection methods.}
\label{tab:detEff}
\begin{tabular}{  c | c  }
	Detection Method & System Detection Efficiency ($\varepsilon_{sys}$) \\ \hline
	PMT (free-space) & $2.171(9)\%$  \\
	PMT (fiber coupled) & $1.777(7)\%$ \\
	SNSPD (fiber coupled) & $4.356(6)\%$ \\
\end{tabular}
\end{table}

\subsection*{\yb Scattering Rates}
The optimized scattering rate of the ion in the \ket{1} state with optimal polarization of the detection beam is given by the expression: \begin{equation}R_{o,opt} = \left(\frac{1}{3}\right) \left(\frac{\Gamma}{2}\right) \left(\frac{s_o}{1 + \frac{2}{3}s_o + \left(\frac{2\Delta}{\Gamma}\right)^2}\right)\label{eq:optScatterRate}\end{equation} where $\Gamma = 2\pi \times 19.6$ MHz is the linewidth of the $^2$P$_{1/2}$ state, $s_o = 2\Omega^2/\Gamma^2$ is the on-resonance saturation parameter with a Rabi frequency $\Omega$, and $\Delta$ is the detuning of the detection beam from the \sfo $\rightarrow$ \pfz cycling transition. The dark pumping rate describes the rate at which the ion will pump to the \ket{0} state after being initialized to the \ket{1} state:\begin{equation}R_d \approx \left(\frac{1}{3}\right) \left(\frac{\Gamma}{2}\right) \left(\frac{2\Omega^2}{\Gamma^2}\right) \left(\frac{\Gamma}{2\Delta_{HFP}}\right)^2 \label{eq:Rd}\end{equation} for which $\Delta_{HFP} = 2\pi \times 2.1$ GHz is the hyperfine splitting of the $^2$P$_{1/2}$ energy level \cite[]{Noek2013}.  The bright pumping rate is the rate at which the ion will off-resonantly pump to the \ket{1} state after initially prepared in the \ket{0} state and is expressed as:  \begin{equation}R_b \approx \left(\frac{2}{3}\right) \left(\frac{\Gamma}{2}\right) \left(\frac{2\Omega^2}{\Gamma^2}\right) \left(\frac{\Gamma}{2(\Delta_{HFP} + \Delta_{HFS})}\right)^2 \label{eq:Rb}\end{equation} where $\Delta_{HFS} = 2\pi \times 12.6$ GHz is the hyperfine splitting of the $^2$S$_{1/2}$ energy level \cite[]{Noek2013}.

\section{Data Availability}
All data from this work is available through the corresponding author upon request. 

\bibliographystyle{naturemag}
\bibliography{references}

\section*{Acknowledgements}
This work was supported by the Army Research Office (ARO) with funds from the IARPA LogiQ program (grant W911NF16-1-0082) and the ARO Quantum Computing Program (grant W911NF15-1-0213), the Office of Naval Research (ONR) MURI programs, and National Aeronautics and Space Administration (NASA) grant on Superdense Teleportation. Part of this work was performed at the Jet Propulsion Laboratory, California Institute of Technology, under contract with the National Aeronautics and Space Administration.

\section*{Author Contributions}
S.C., C.C., G.V., and J.K. conceived and carried out the state detection and crosstalk characterization experiments and analysis. S.C. and J.K. prepared the manuscript. E.E.W., M.D.S., V.B.V. and S.W.N contributed to the design and fabrication of the SNSPD devices used in these experiments. 

\section*{Competing Interests}
The authors declare no competing interests.
\end{document}